\documentclass[preprint,superscriptaddress,prb]{revtex4}

\usepackage{graphicx}
\usepackage{epstopdf}
\usepackage{bbm}
\usepackage{amsmath}
\usepackage{eqnarray}

\begin{document}

\title{Dielectric permittivity, conductivity and breakdown field of hexagonal boron nitride}

\author{A. Pierret}\email{aurelie.pierret@phys.ens.fr}
\affiliation{Laboratoire de Physique de l'Ecole normale sup\'erieure, ENS, Universit\'e
PSL, CNRS, Sorbonne Universit\'e, Universit\'e de Paris, 24 rue Lhomond, 75005 Paris, France}
\author{D. Mele}
\affiliation{Laboratoire de Physique de l'Ecole normale sup\'erieure, ENS, Universit\'e
PSL, CNRS, Sorbonne Universit\'e, Universit\'e de Paris, 24 rue Lhomond, 75005 Paris, France}
\author{H. Graef}\affiliation{Laboratoire de Physique de l'Ecole normale sup\'erieure, ENS, Universit\'e
PSL, CNRS, Sorbonne Universit\'e, Universit\'e de Paris, 24 rue Lhomond, 75005 Paris, France}
\author{J. Palomo}\affiliation{Laboratoire de Physique de l'Ecole normale sup\'erieure, ENS, Universit\'e
PSL, CNRS, Sorbonne Universit\'e, Universit\'e de Paris, 24 rue Lhomond, 75005 Paris, France}
\author{T. Taniguchi} \affiliation{Advanced Materials Laboratory, National Institute for Materials Science, Tsukuba, Ibaraki 305-0047,  Japan}
\author{K. Watanabe} \affiliation{Advanced Materials Laboratory, National Institute for Materials Science, Tsukuba, Ibaraki 305-0047, Japan}
\author{Y. Li}\affiliation{Laboratoire des Multimat\'eriaux et Interfaces, UMR CNRS 5615, Univ Lyon, Universit\'e Claude Bernard Lyon 1,
F-69622 Villeurbanne, France}
\author{B. Toury}\affiliation{Laboratoire des Multimat\'eriaux et Interfaces, UMR CNRS 5615, Univ Lyon, Universit\'e Claude Bernard Lyon 1,
F-69622 Villeurbanne, France}
\author{C. Journet}\affiliation{Laboratoire des Multimat\'eriaux et Interfaces, UMR CNRS 5615, Univ Lyon, Universit\'e Claude Bernard Lyon 1,
F-69622 Villeurbanne, France}
\author{P. Steyer}\affiliation{Universit\'e de Lyon, MATEIS, UMR CNRS 5510, INSA-Lyon, F-69621 Villeurbanne cedex, France}
\author{V. Garnier}\affiliation{Universit\'e de Lyon, MATEIS, UMR CNRS 5510, INSA-Lyon, F-69621 Villeurbanne cedex, France}
\author{A. Loiseau}\affiliation{Laboratoire d'Etude des Microstructures (LEM), ONERA, CNRS, Universit\'e Paris-Saclay, 92322 Ch\^atillon, France}
\author{J-M. Berroir}\affiliation{Laboratoire de Physique de l'Ecole normale sup\'erieure, ENS, Universit\'e
PSL, CNRS, Sorbonne Universit\'e, Universit\'e de Paris, 24 rue Lhomond, 75005 Paris, France}
\author{E. Bocquillon}\affiliation{Laboratoire de Physique de l'Ecole normale sup\'erieure, ENS, Universit\'e
PSL, CNRS, Sorbonne Universit\'e, Universit\'e de Paris, 24 rue Lhomond, 75005 Paris, France}
\affiliation{Physikalisches Institut, Universit\"at zu K\"oln, Z\"ulpicher Strasse 77, 50937 K\"oln}
\author{G. F\`eve}\affiliation{Laboratoire de Physique de l'Ecole normale sup\'erieure, ENS, Universit\'e
PSL, CNRS, Sorbonne Universit\'e, Universit\'e de Paris, 24 rue Lhomond, 75005 Paris, France}
\author{C. Voisin}
\affiliation{Laboratoire de Physique de l'Ecole normale sup\'erieure, ENS, Universit\'e
PSL, CNRS, Sorbonne Universit\'e, Universit\'e de Paris, 24 rue Lhomond, 75005 Paris, France}
\author{E. Baudin} \affiliation{Laboratoire de Physique de l'Ecole normale sup\'erieure, ENS, Universit\'e
PSL, CNRS, Sorbonne Universit\'e, Universit\'e de Paris, 24 rue Lhomond, 75005 Paris, France}
\author{M. Rosticher}\affiliation{Laboratoire de Physique de l'Ecole normale sup\'erieure, ENS, Universit\'e
PSL, CNRS, Sorbonne Universit\'e, Universit\'e de Paris, 24 rue Lhomond, 75005 Paris, France}
\author{B. Pla\c{c}ais} \email{bernard.placais@phys.ens.fr}
\affiliation{Laboratoire de Physique de l'Ecole normale sup\'erieure, ENS, Universit\'e
PSL, CNRS, Sorbonne Universit\'e, Universit\'e de Paris, 24 rue Lhomond, 75005 Paris, France}

\begin{abstract}
In view of the extensive use of hexagonal boron nitride (hBN) in 2D material electronics, it becomes important to refine its dielectric characterization in terms of low-field permittivity and high-field strength and conductivity up to the breakdown voltage. The present study aims at filling this gap using DC and RF transport in two Au-hBN-Au capacitor series of variable thickness in the $10$--$100\;\mathrm{nm}$ range, made of large  high-pressure, high-temperature (HPHT) crystals and a polymer derivative ceramics (PDC) crystals. We deduce an out-of-plane low field dielectric constant $\epsilon_\parallel=3.4\pm0.2$ consistent with the theoretical prediction of Ohba et al., that narrows down the generally accepted window $\epsilon_\parallel=3$--$4$. The DC-current leakage at high-field is found to obey the Frenkel-Pool law for thermally-activated trap-assisted electron transport with  a dynamic dielectric constant $\epsilon_\parallel\simeq3.1$ and a trap energy $\Phi_B\simeq1.3\;\mathrm{eV}$, that is comparable with standard technologically relevant dielectrics.
\end{abstract}

\maketitle

\section{Introduction}

Hexagonal boron nitride (hBN) is a van der Waals crystal insulator introduced in graphene electronics a decade ago [\onlinecite{Dean2010nnano}] and since then extensively used as encapsulant [\onlinecite{Mayorov2011nl}], tunnel barrier [\onlinecite{Lee2011apl,Britnell2012nl,Kim2020nelec}], or gate dielectric in 2D material electronics  [\onlinecite{Novoselov2016science,Illarionov2020ncomm}]. In view of its technological relevance, it is important to improve its characterization both in terms of low-field permittivity and  high-field dielectric conductivity and breakdown field. Accepted values for the DC dielectric permittivity constant lie in the broad range $\epsilon^\parallel=3$--$4$ [\onlinecite{Dean2010nnano,Yang2021prl,Hattori2016acsami,Veyrat2019nl}], whereas breakdown fields are more scattered, with $E_{BD}=4$--$10\;\mathrm{MV/cm}$ [\onlinecite{Hattori2016acsami,Ahmed2018afm}], depending on material quality and breakdown-field definition criterion. The purpose of the present study, which is based on DC and RF transport in Au-hBN-Au capacitors made of two types of hBN crystals, is to narrow down the uncertainty in permittivity, to shed light onto the dielectric breakdown mechanism, and to use these characterizations to benchmark the two hBN crystal sources.

We have used two series of Au-hBN-Au capacitors made of large exfoliated hBN crystals. Crystals are grown either under high-pressure high-temperature (HPHT samples) as described in Ref.[\onlinecite{Taniguchi2007jcg}], or with a polymer derivative ceramics (PDC) route described in Ref.[\onlinecite{Li2020acs-anm}]. Exfoliated hBN flakes, of  thickness $d\sim10$--$100\;\mathrm{nm}$, are sandwiched between Au electrodes of lateral dimensions  $L\times W=10\times10\;\mathrm{\mu m}$. A significant fraction of samples ($17$ HPHTs and $11$ PDCs of the total $41$ capacitors) follows the parallel-plate capacitance law with a dielectric constant $\epsilon^\parallel\simeq3.4\pm0.2$. The other $14$ samples deviate from from this law with  lower capacitance values, presumably due to process imperfections involving spurious air gaps,  due to dust or bubbles between the hBN flake and the bottom electrode. Besides, we do not see significant differences between HPHT and PDC crystals in terms of permittivity.

The dielectric strength is characterized by monitoring the leakage current at high bias which is analyzed in terms of a bulk conductivity. This analysis is carried out on a subset of $7$ capacitors having survived the high-bias training. As shown in Ref.[\onlinecite{Hattori2016acsami}], dielectric conductivity is a three step process, starting by a threshold-less exponential current growth, followed by a quasi-saturation at a breakdown field $E_{BD}$ and culminating by an irreversible current runaway for $E\gtrsim E_{BD}$ usually leading to sample breakdown. Here we focus on the pre-breakdown regime $E\lesssim E_{BD}$ where moderate current densities ($J< J_{BD}\simeq 0.5\;\mathrm{A/cm^2}$) are applied that secure device integrity. In these conditions, we observe a bias-reversible, reproducible and polarity-independent behavior. The leakage current obeys a standard exponential growth with voltage which precludes unambiguous determination of a dielectric breakdown voltage (see Fig.\ref{figure2}-a). However, when using an arbitrary  breakdown current criterion $J\lesssim J_{BD}$, we find an overall increase of the breakdown voltage with thickness which suggests the relevance of a breakdown field, i.e. a bulk scenario, and justifies our breakdown analysis in terms of conductivity, e.g. $J/E(E)$ below a breakdown conductivity $\sigma_{BD}\sim 1\;\mathrm{\mu\Omega/cm}$.  We find that the conductivity, $\sigma=J/E$ obeys the Frenkel-Pool (FP) law (Eq.(\ref{Frenkel-Pool}) below), corresponding to a trap-assisted, thermally-activated, bulk Schottky transport [\onlinecite{Frenkel1938pr,Sze2007wiley}]. Its signature lies in the linear dependence, $\ln{\sigma/\sigma_{BD}}=f(\sqrt{E/\epsilon^\parallel T^2})$ (see Fig.\ref{figure2}-b), observed in the $[10^{-4}$--$10^{-1}]\;\mathrm{\mu S/cm}$ range at room temperature ($T=300\;\mathrm{K}$). The FP activation scenario is confirmed by the temperature dependence measured in one representative sample (inset of Fig.\ref{figure2}-b). The robustness of the field dependence contrasts with the large variability (within a factor $10^6$) of the conductivity prefactor. We assign the latter to a variability in the deep-level donor energy $\Phi_B\simeq0.9$--$1.3\;\mathrm{eV}$, with a logarithmical precision in the $\sigma_{BD}$ prefactor. Remarkably, five devices show very similar Frenkel-Pool plots with deep-level traps energy  $\Phi_{B,hBN}\simeq1.27\pm0.03\;\mathrm{eV}$, suggesting the existence of a quasi-intrinsic limit. This value is quite typical of that of technology-relevant insulators, such as Si$_3$N$_4$ where $\Phi_{B}\simeq1.3\pm0.2\;\mathrm{eV}$  [\onlinecite{Sze1967apl}], or SiO$_2$ where $\Phi_{B}\simeq1\;\mathrm{eV}$ [\onlinecite{Harrel1999tsf}] are reported. Some PDC-grown devices show however traps with a smaller energy, which opens routes for improvement of this less mature growth process.

\section{Capacitor fabrication and setup}\label{setup}

\begin{figure}[h!!!!]
\centerline{\includegraphics[width=17cm]{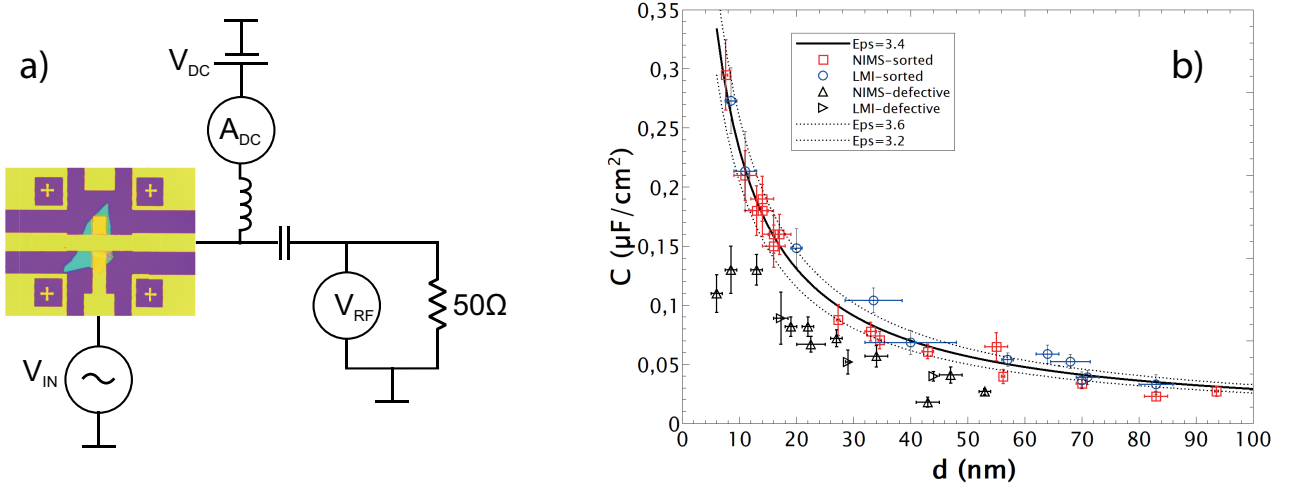}}
\caption{Dielectric constant deduced from a series of Au-hBN-Au capacitors of thickness $d=6$--$98\;\mathrm{nm}$
and lateral dimensions $L\times W=10\times10\;\mathrm{\mu m}$. Panel a) : scheme of the measuring setup. Panel b): capacitance of the series of $41$ devices. A subset of $17$ HPHT-type and $10$ PDC-type capacitors follow the parallel-plate formula $C=\epsilon^\parallel\epsilon_0 LW/(d+\epsilon^\parallel \delta)$, where $\delta\simeq1\;\mathrm{nm}$ accounts for air-gap contribution associated with bottom Au-plate roughness  $2\delta$. From this subset we can refine the precision of the dielectric constant at  $\epsilon^\parallel=3.4\pm0.2$.}
 \label{figure1}
\end{figure}

We have fabricated $41$ Au-hBN-Au capacitors, $27$ of them with  HPHT-hBN from  NIMS  and $14$ with the PDC-hBN from  LMI. The growth technics, as well as structural and optical characterizations, of these high-quality crystals are detailed in [\onlinecite{Taniguchi2007jcg}]  and  [\onlinecite{Li2020acs-anm,Maestre2022preprint}] respectively. The capacitors were deposited on high-resistivity Si substrates, suitable for RF measurement, that are covered with a $285\;\mathrm{nm}$-thick SiO2 layer. The bottom electrode and the RF coplanar waveguide structure are first deposited using laser lithography and thermal evaporation. The bottom $110/5\;\mathrm{nm}$  Au/Cr electrode is buried into  SiO2 to ensure a planar surface for hBN-flakes transfer. Planarisation is completed by mechanically polishing the small metallic pitches at the gold edges with isopropyl alcohol (IPA). This process minimizes air gaps between hBN and the bottom gold electrode which are ultimately limited by Au roughness of amplitude $2\delta\simeq 1.5\;\mathrm{nm}$, as measured by atomic force microscopy (AFM). hBN crystals were mechanically exfoliated with polydimethylsiloxane (PDMS). We use poly-propylene carbonate PPC-PDMS stamps to dry transfer the hBN flakes on the bottom electrode. A second lithography step allows covering the hBN flake with a second Au/Cr electrode, which is conformal to the hBN dielectric and therefore airgap-free. The obtained structures form a nominally $10\times10\;\mathrm{\mu m}$ capacitor in a parallel-plate configuration.  After annealing ($1\mathrm{h}$ at $240\;\mathrm{Celsius}$ under N$_2$) the  capacitor's hBN thickness is determined by AFM and falls in the range $d=6$--$98\;\mathrm{nm}$. This range exceeds the minimal thickness ($10$-$20\;\mathrm{nm}$) for mobility-preserving encapsulation, is relevant for gate dielectric applications, and reaches the value ($\sim100\;\mathrm{nm}$) for fully-developed radiative cooling [\onlinecite{Yang2018nnano,Baudin2020adfm}]. It is however not relevant for tunnel-barrier applications which are described elsewhere [\onlinecite{Lee2011apl,Britnell2012nl,Kim2020nelec}]. 
The lateral size of capacitors was deliberately maximized at the limits of our exfoliation technique to increase experimental resolution, minimize spurious edge effects, and address the homogenous properties of hBN. It is precisely measured by scanning electron microscopy. We have indifferently used colinear and perpendicular source and drain electrodes, the later geometry being shown in the optical image of Fig.\ref{figure1}-a.

High frequency admittance measurements were carried out in a Janis (cryogenic) probe station under vacuum  at room temperature (see sketch in Fig.\ref{figure1}-a). The two-port scattering parameters $S_{ij}$ of the capacitor were measured using an Anritsu MS4644B vectorial network analyzer (VNA) in the $10\;\mathrm{MHz}$--$10\;\mathrm{GHz}$ range.
A short-open-load-reciprocal protocol was used to calibrate the wave propagation until the probe tips. As explained in Ref.[\onlinecite{Graef2018jpm}], the wave propagation in the coplanar access of the capacitor is  de-embedded by calculating the ABCD matrix from the S parameters of a symmetric thruline reference structure. To correct for residual  parasitic stray capacitance effects, we convert the previous ABCD matrices into complex admittance matrices and subtract the contribution of a dummy reference structure of same geometry but devoid of the central Au-hBN-Au capacitor. The total capacitance is directly deduced from the low frequency (sub-GHz) imaginary part $Y_{12} = j \omega C$.

Breakdown measurements have been carried out on the same setup by monitoring the leakage current as function of the applied DC voltage (Keithley 2400 Source-Measure Unit). The same procedure has been applied at low temperature on a representative capacitor to check the activation mechanism at stake in the dielectric breakdown.

\section{Dielectric constant}\label{dielectric-constant}

Fig.\ref{figure1}-b shows capacitances of the full set of $41$ tested capacitors versus hBN thickness. Capacitance data, deduced from  VNA measurements, are first compared with complementary data obtained by sub-MHz Lock-In techniques (not shown) to ascertain their frequency independence. Data are scattered, but we can still find a significant fraction of devices, of both HPHT and PDC sources, showing capacitance-data accumulation along an upper limit given by the parallel-plate capacitance formula $C=\epsilon_0\epsilon^\parallel LW/d^*$ (red and blue symbols in Fig.\ref{figure1}-b), where $\epsilon^\parallel\simeq3.4\pm0.2$ and $d^*=d+\epsilon^\parallel\delta$ is an effective dielectric thickness accounting for spurious air-gap contributions due to (bottom) metal roughness $2\delta\simeq2\;\mathrm{nm}$. One third of the series ($14$ devices, black symbols in Fig.\ref{figure1}-b), exhibiting lower capacitance values caused by process imperfections including spurious air gaps, presumably due to dust or bubbles between the hBN flake and the bottom electrode, are discarded. From the selected devices ($17$ HPHTs and $11$ PDCs), we are able to narrow down the dielectric constant window and provide a recommended value of the hBN dielectric constant $\epsilon^\parallel=3.4\pm0.2$. This value exceeds by $13\%$ the $\epsilon^\parallel\simeq 3.0$ reported in metal-hBN-graphene quantum-Hall devices [\onlinecite{Yang2018nnano,Yang2021prl}]. This apparent disagreement can be lifted when considering the series quantum capacitance of the graphene electrode in these thin hBN samples [\onlinecite{Pallecchi2011prb,Yang2021prl}], as well as roughness-induced air-gaps at the bottom electrodes. Our measured permittivity turns out to be consistent with the theoretical prediction $\epsilon_{\parallel}=3.38$  of Ref.[\onlinecite{Ohba2001prb}], following an ab-initio approach which already quantitatively predicts correctly the optical permittivity and its anisotropy [\onlinecite{Segura2018prm}].

 \begin{figure}[h!!!!]
\centerline{\includegraphics[width=17cm]{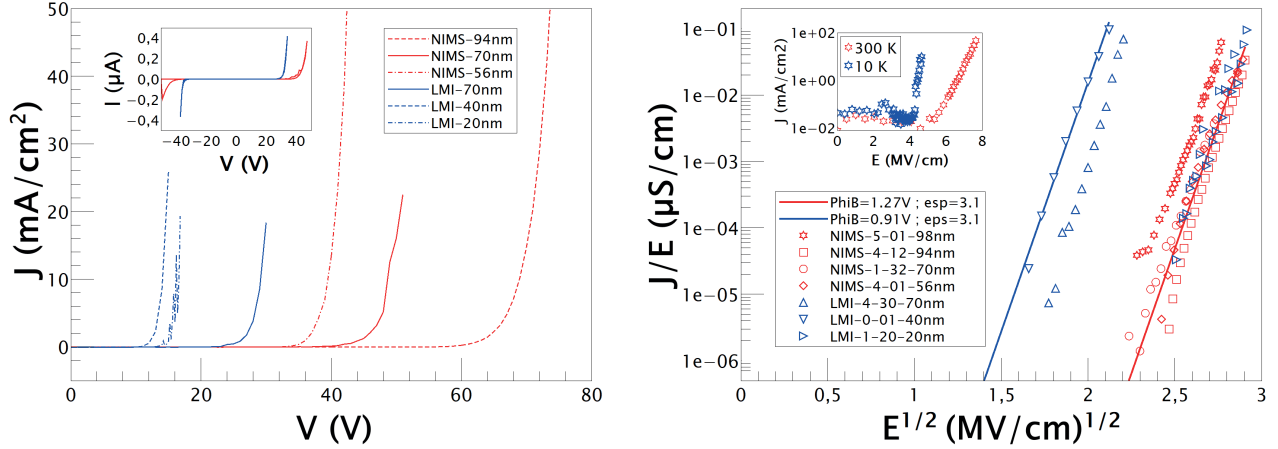}}
\caption{Room temperature dielectric breakdown in HPHT (red symbols/lines) and PDC (blue symbols/lines) hBN capacitors of $10\times10\;\mathrm{\mu m}$ lateral dimension.
Panel a): exponential growth of leakage current at large voltage. Inset shows bipolar characteristics over a broader current density range. Pre-breakdown voltage increases with hBN thickness and is typically larger
in HPHT-based devices than in PDC-based ones.
Panel b): Frenkel-Pool plot of the high-field hBN conductivity as function of electric field, showing a scaling of the field-dependence.
The dielectric strength (or breakdown field) takes a maximum shared by 4 HPHT samples and 1 PDC sample.
It obeys the Frenkel-Pool law (red line) in Eq.(\ref{Frenkel-Pool}) with $\Phi_B=1.27\pm0.03\;\mathrm{V}$,
 taking $\sigma_{BD}=0.1\;\mathrm{\mu S/cm}$ as an estimate of the fully ionized donor conductivity, and
 $\epsilon_{BD}=3.1$. Two PDC samples (upward- and downward-pointing blue triangles) show the same Frenkel-Pool field dependence (blue line)
 however with smaller trap energies  $\Phi_B=0.9$ and $1.0\;\mathrm{V}$. Inset shows the temperature dependence of high-field current, measured on a representative sample. The increase of the slope at 10K illustrates the thermally activated nature of transport.}
 \label{figure2}
\end{figure}

\section{Breakdown current and conductivity}\label{conductivity}

Figure \ref{figure2}-a shows typical current-voltage characteristics of capacitors showing a strongly non-linear onset of current.
Characteristics become irreversible at large current density, $J\sim0.1\;\mathrm{\mu A/cm^2}$,
eventually leading to sample breakdown for $J\gtrsim J_{BD}\gtrsim1\;\mathrm{\mu A/cm^2}$. At the lower bias of Fig.\ref{figure2}-a,  I-V curves are reproducible and bias symmetric (inset of Fig.\ref{figure2}-a). As seen in the figure, the breakdown voltage $V_{BD}$ shows a tendency  to increase with hBN thickness $d$, suggesting a bulk origin of breakdown.

A deeper insight into the breakdown mechanism is provided by Fig.\ref{figure2}-b which reveals a $\ln(J/E)\propto\sqrt{E}$
scaling of conductivity. It corresponds to the Frenkel-Pool (FP) effect [\onlinecite{Frenkel1938pr,Sze1967apl,Sze2007wiley}],
where a finite dielectric conductivity $J/E$ arises due to thermal de-trapping of deep-level electrons of energy $e\Phi_B\gg k_BT$.
Conductivity increases at large voltage as electric field lowers the barrier height $\Phi_B$ by an amount
$\sqrt{eE/\pi\epsilon_{BD}\epsilon_{0}}$. At ultimate fields, it eventually leads to full ionization of traps whenever
$(\Phi_B-\sqrt{eE/\pi\epsilon_{BD}\epsilon_{0}})\lesssim k_BT$, which defines a breakdown electric field
$E_{BD}$. In this picture, $E_{BD}=\Phi_B/r_0$ is on the order of the impurity field, where $r_0=e/\pi\epsilon_{BD}\epsilon_{0}\Phi_B$ is the screening length. The FP conductivity writes [\onlinecite{Sze2007wiley}]:
\begin{equation}
\frac{J}{E}=\sigma_{BD}\times \exp\left[-e\frac{\Phi_B-\sqrt{eE/\pi\epsilon_{BD}\epsilon_{0}}}{k_BT}\right]\quad ,
\label{Frenkel-Pool}\end{equation}
where the breakdown conductivity $\sigma_{BD}=J_{BD}/E_{BD}\simeq N_Te\mu$ corresponds to the band conductivity (mobility $\mu$) for fully ionized traps (concentration $N_T$). Eq.(\ref{Frenkel-Pool}) holds in the field range $E_{FN}\lesssim E \lesssim E_{BD}$, or $J=10^{-4}$--$10^{-1}\;\mathrm{\mu A/cm^2}$, where $E_{FN}$ is the Fowler-Nordheim tunneling limit for a defect-free thin triangular barrier [\onlinecite{Sze1967apl}].

As seen in Fig.\ref{figure2}-b, the field dependence of breakdown conductivity does obey the FP mechanism for both hBN sources, with a slope
solely determined by $\epsilon_{BD}$. Data fitting suggests a small field-suppression of permittivity with  $\epsilon_{BD}\simeq3.1<\epsilon^\parallel\simeq3.4$. In contrast to the universality of slope, the prefactor $\sigma_{BD} \exp\left(-\frac{e\Phi_B}{k_BT}\right)$ shows sample variability, exhibiting variations by six orders of magnitude. This observation highlights the strong sensitivity of breakdown to material quality which, in the FP scenario, mainly stems from the variability in the trap potential $\Phi_B$ (and $E_{BD}$),  as that of $N_T$ and $\mu$ in $\sigma_{BD}$ cannot explain such a large scatter alone.  For a quantitative estimation of $\Phi_B$ we set  $\sigma_{BD}\sim0.1\;\mathrm{\mu S/cm}$, corresponding to  $J_{BD}\sim 0.5\;\mathrm{A/cm^2}$ and $E_{BD}\sim 5\;\mathrm{MV/cm}$. This breakdown current  corresponds is typical of quasi saturation values observed in our samples, and in the literature (Fig.5 in Ref.[\onlinecite{Hattori2016acsami}]). With this procedure, we extract  $\Phi_B=0.9$--$1.3\;\mathrm{V}$ in Fig.\ref{figure2}-b. Remarkably, we observe an accumulation of conductivity data (4 HPHT and 1 PDC capacitors) along an upper limit represented by the $\Phi_B=1.27\pm0.03\;\mathrm{V}$ line. This suggests the existence of a dielectric strength limit in high-quality hBN crystals. Dispersion among that data-subset of highest-quality samples corresponds to $\Delta\Phi_B=0.06\;\mathrm{V}$, or a variation of $\sigma_{BD}$ in the range $0.01$--$1\;\mathrm{\mu S/cm}$. Taking a typical insulator mobility $\mu\lesssim1\;\mathrm{cm^2/Vs}$, this translates into a trap density  $N_T\sim 10^{12}$--$10^{13}\;\mathrm{cm^{-3}}$ and a trap number $N_T d L W\sim 10$--$100$.

To further establish the FP mechanism of breakdown we have added, in the insert of Fig.\ref{figure2}-b, a comparison between $10\;\mathrm{K}$ and $300\;\mathrm{K}$ breakdown current measurements performed on an additional sample (NIMS-5-01-98 $\;\mathrm{nm}$), which illustrates the strong temperature dependence of activated FP transport, at variance with tunneling-based scenarios.

\section{Conclusion} \label{conclusion}

Our estimate of hBN dielectric permittivity, $\epsilon^\parallel\equiv\epsilon_0^{\parallel c}=3.4\pm0.2$, agrees with the calculations of Ohba et al., which predicts :  $\epsilon_\infty^{\perp c}=4.85$,  $\epsilon_\infty^{\parallel c}=2.84$, $\epsilon_0^{\perp c}=6.61$,  $\epsilon_0^{\parallel c}=3.38$ [\onlinecite{Ohba2001prb}]. This completes previous results based on optical measurements giving $\epsilon_\infty^{\perp c}=4.95$,  $\epsilon_\infty^{\parallel c}=2.86$, $\epsilon_0^{\perp c}=6.96$ [\onlinecite{Segura2018prm}]. The excellent agreement in the four relevant dielectric constants  of hBN gives strong confidence in the ab-initio technique to provide reliable predictions of static and dynamical properties of BN crystals including those of the zinc-blende and wurtzite crystals.

The relevance of the 3D Frenkel-Pool mechanism of conductivity in the 2D hBN was not granted as 2D materials may sustain specific mechanisms. We demonstrate here that it works for c-axis transport, but the situation can be different for in-plane electric fields with the opening of new leakage channels associated with charges gliding in-between hBN planes. The identification of deep-level traps responsible for breakdown is beyond the scope of our work, especially as leakage current alone cannot identify the acceptor/donor nature of the levels. We find $\Phi_{B}\simeq1.27\pm0.03\;\mathrm{eV}$ in hBN, which is larger than the $1\;\mathrm{eV}$ reported in SiO$_2$ [\onlinecite{Harrel1999tsf}], and comparable with the  $1.3\pm0.2\;\mathrm{eV}$ in Si$_3$N$_4$ [\onlinecite{Sze1967apl}]. In the latter case, deep traps are attributed to silicon-dangling-bond centers [\onlinecite{Krick1988prb}]. This trap energy determines a maximum breakdown field $E_{BD}\simeq 5\;\mathrm{MV/cm}$, defined as the threshold for current quasi-saturation.

Finally we conclude on the strong similarity of HTHP and PDC hBN crystals in terms of dielectric permittivity and strength, with however a better yield in terms of intrinsic dielectric breakdown for the NIMS samples, which can to a large extent be attributed to a longer maturity of the growth technique.

\begin{acknowledgments}
 The research leading to these results has received partial funding from the the European Union ``Horizon 2020'' research and innovation programme under grant agreement No.881603 "Graphene Core 3", the ANR-14-CE08-018-05 "GoBN" and  ANR-21-CE24-0025-01 "ELuSeM".
\end{acknowledgments}


\begin{thebibliography}{}


\bibitem{Dean2010nnano}
C. R. Dean, A. F. Young, I.Meric, C. Lee, L. Wang, S. Sorgenfrei, K. Watanabe, T. Taniguchi, P. Kim, K. L. Shepard and J. Hone,
\emph{Nature Nanotechnol.} \textbf{5}, 722 (2010). \emph{Boron nitride substrates for high-quality graphene electronics}

\bibitem{Mayorov2011nl}
A.S. Mayorov, R.V. Gorbachev, S.V. Morozov, L. Britnell, R. Jalil, L.A. Ponomarenko, P. Blake, N.S. Novoselov, K. Watanabe, T. Taniguchi, A. Geim,
Nano Lett. 11, 2396 (2011).
\emph{ Micrometer-Scale Ballistic Transport in Encapsulated Graphene at Room Temperature}

\bibitem{Lee2011apl}
G-H. Lee, Y-J. Yu, C. Lee, C. Dean, K. L. Shepard, P. Kim, and J. Hone,
\emph{Appl. Phys. Lett.} \textbf{99}, 243114 (2011). \emph{Electron tunneling through atomically flat and ultrathin hexagonal boron nitride}

\bibitem{Britnell2012nl}
L. Britnell, R.V. Gorbachev, B.D. Belle, F. Schedin, M.I. Katsnelson, L. Eaves, S.V. Morozov, A.S. Mayorov, N.M.R. Peres, A.H. Castro Neto, J. Leist, A.K. Geim, L.A. Ponomarenko, K.S. Novoselov,  \emph{Nano Lett.} \textbf{12}, 1707 (2012).
\emph{Electron Tunneling through Ultrathin Boron Nitride Crystalline Barriers}


\bibitem{Kim2020nelec}
M. Kim, E. Pallecchi, R. Ge, X. Wu, G. Ducournau, J.C. Lee, H. Happy,  D. Akinwande,
\emph{Nature Electronics}  \textbf{3},  479 (2020).
\emph{Analogue switches made from boron nitride monolayers for application in 5G and terahertz communication systems}


\bibitem{Novoselov2016science}
K. S. Novoselov, A. Mishchenko, A. Carvalho, A. H. Castro Neto, \emph{Science} \textbf{353}, 461 (2016).
\emph{2D materials and van der Waals heterostructures}

\bibitem{Illarionov2020ncomm}
Y. Y. Illarionov, T. Knobloch, M. Jech, M. Lanza, D. Akinwande, M. I. Vexler, T. Mueller, M. C. Lemme, G. Fiori, F. Schwierz,  T. Grasser, \emph{Nat. Commun.} \textbf{11}, 3385 (2020).
\emph{Insulators for 2D nanoelectronics: the gap to bridge}

\bibitem{Veyrat2019nl}
L. Veyrat, A. Jordan, K. Zimmermann, F. Gay, K. Watanabe, T. Taniguchi, H. Sellier, and B. Sac\'ep\'e,
\emph{Nano Lett. } \textbf{19},  635 (2019)
\emph{Low-Magnetic-Field Regime of a Gate-Defined Constriction in High-Mobility Graphene}

\bibitem{Hattori2016acsami}
Y. Hattori, T. Taniguchi, K. Watanabe, and Kosuke Nagashio, \emph{ACS Appl. Mater. Interfaces} \textbf{8}, 27877 (2016).
\emph{Anisotropic Dielectric Breakdown Strength of Single Crystal Hexagonal Boron Nitride}

\bibitem{Yang2021prl}
F. Yang, A.A. Zibrov, R. Bai, T. Taniguchi, K. Watanabe, M.P. Zaletel, and A. F. Young,
\emph{Phys. Rev. Lett. }\textbf{126}, 156802 (2021)
\emph{Experimental Determination of the Energy per Particle in Partially Filled Landau Levels}

\bibitem{Ahmed2018afm}
F. Ahmed, S. Heo, Z. Yang, F. Ali, C. H. Ra, H-I. Lee, T. Taniguchi, J. Hone, B. H. Lee, and W. J. Yoo, \emph{Adv. Funct. Mater.} \textbf{28}, 1804235 (2018).
\emph{Dielectric Dispersion and High Field Response of Multilayer Hexagonal Boron Nitride}

\bibitem{Taniguchi2007jcg}
T. Taniguchi, K. Watanabe, Journal of Crystal Growth \textbf{303}, 525 (2007).
\emph{Synthesis of high-purity boron nitride single crystals under high pressure by using Ba-BN solvent}

\bibitem{Li2020acs-anm}
Y. Li, V. Garnier, P. Steyer, C. Journet,  B. Toury, \emph{ACS Appl. Nano Mater.} \textbf{3}, 2, 1508 (2020).
\emph{Millimeter-Scale Hexagonal Boron Nitride Single Crystals for Nanosheet Generation}

\bibitem{Maestre2022preprint}
 C. Maestre, Y. Li, V. Garnier, P. Steyer, S. Roux, A. Plaud, A. Loiseau, J. Barjon, L. Ren, C. Robert, B Han, X. Marie, C. Journet, B. Toury, arXiv:2201.07673 (2022).
\emph{From the synthesis of hBN crystals to their use as nanosheets for optoelectronic devices}

\bibitem{Ohba2001prb}
N. Ohba, K. Miwa, N. Nagasako,  A. Fukumoto, \emph{Phys. Rev. B} \textbf{63}, 115207 (2001).
\emph{First-principles study on structural, dielectric, and dynamical properties for three BN polytypes}

\bibitem{Yang2018nnano}
W. Yang, S. Berthou, X. Lu, Q. Wilmart, A. Denis, M. Rosticher, T. Taniguchi, K. Watanabe, G. F\`eve, J.M. Berroir, G. Zhang,
C. Voisin, E. Baudin, B. Pla\c{c}ais, \emph{Nature Nanotechnol.} \textbf{13}, 47 (2018).
\emph{A graphene Zener-Klein transistor cooled by a hyperbolic substrate}

\bibitem{Baudin2020adfm}
E. Baudin, C. Voisin,  B. Pla\c{c}ais, \emph{Adv. Funct. Mater.}, 1904783 (2019).
\emph{Hyperbolic Phonon Polariton Electroluminescence as an Electronic Cooling Pathway}

\bibitem{Pallecchi2011prb}
E. Pallecchi, A.C. Betz, J. Chaste, G. F\`eve, B. Huard, T. Kontos, J.-M. Berroir and B. Pla\c{c}ais,
\emph{Phys. Rev.  B} \textbf{83}, 125408 (2011). \emph{Transport scattering time probed through rf admittance of a graphene capacitor}

\bibitem{Frenkel1938pr}
J. Frenkel, \emph{Phys. Rev.} \textbf{54}, 647 (1938).
\emph{On Pre-Breakdown Phenomena in Insulators and Electronic Semi-Conductors}

\bibitem{Sze2007wiley}
S.M. Sze and K. Ng, \emph{Physics of Semicondcutor Devices,  Wiley-2007-3rd edition}, Section 6.7.2.

\bibitem{Sze1967apl}
S.M. Sze, \emph{J. Appl. Phys.} \textbf{38}, 2951 (1967).
\emph{Current Transport and Maximum Dielectric Strength of Silicon Nitride Films}

\bibitem{Harrel1999tsf}
W.R. Harrell, J. Frey, \emph{\emph{Thin Solid Films}} \textbf{352}, 195 (1999).
\emph{Observation of Poole Frenkel effect saturation in SiO2 and other insulating films}

\bibitem{Graef2018jpm}
H. Graef, D. Mele, M. Rosticher, C. Stampfer, T. Taniguchi, K. Watanabe, E. Bocquillon,  G. F\`eve, J.M. Berroir, E.T.H. Teo,  B. Pla\c{c}ais,  \emph{J. Phys. Mater.} \textbf{1},  01LT02 (2018).
\emph{Ultra-long wavelength Dirac plasmons in graphene capacitors.}

\bibitem{Pelini2019prm}
T. Pelini, C. Elias, R. Page, L. Xue, S. Liu, J. Li, J. H. Edgar, A. Dr\'eau, V. Jacques, P. Valvin, B. Gil,  G. Cassabois, \emph{Phys. Rev. Mat.}
\textbf{3}, 094001 (2019).
\emph{Shallow and deep levels in carbon-doped hexagonal boron nitride crystals}

\bibitem{Segura2018prm}
A. Segura, L. Artus, R. Cusco, T. Taniguchi, G. Cassabois, and B. Gil, \emph{Phys. Rev. Mat. B} \textbf{2}, 024001 (2018).
\emph{Natural optical anisotropy of h-BN: Highest giant birefringence in a bulk crystal through the mid-infrared to ultraviolet range}

\bibitem{Krick1988prb}
D. T. Krick, P. M. Lenahan, and J. Kanicki,\emph{ Phys. Rev. B} \textbf{38}, 8226 (1988).
\emph{Nature of the dominant deep trap in amorphous silicon nitride}

\end{thebibliography}
\end{document}